\documentclass[amsmath,amssymb, aps, bibtex, prl,letter,reprint,
 shortbibliography]{revtex4-2}
\usepackage{verbatim}
\usepackage[english]{babel}
\usepackage{hyperref} % Hyperlinks for references. 
\usepackage{svg}
\usepackage{graphicx}% Include figure files
\usepackage{dcolumn}% Align table columns on decimal point
\usepackage{bm}% bold math
\usepackage{amsmath}
\usepackage{enumerate}
\usepackage[normalem]{ulem}
\usepackage{float}
\usepackage{refcount}
\newcounter{SIfootnote}

\newcommand{\be}{\begin{equation}}
\newcommand{\ee}{\end{equation}}
\newcommand{\bae}{\begin{eqnarray}}
\newcommand{\eae}{\end{eqnarray}}

\begin{document}

\preprint{arXiv}

\title{Robust Scaling in Human Brain Dynamics Despite Latent Variables and Limited Sampling Distortions}

\author{Rub\'en Calvo} 
\affiliation{Departamento de Electromagnetismo y F{\'\i}sica de la Materia and Instituto Carlos I
de F{\'\i}sica Te{\'o}rica y Computacional. Universidad de Granada.E-18071, Granada, Spain}
 \author{Carles Martorell}
\affiliation{Departamento de Electromagnetismo y F{\'\i}sica de la Materia and Instituto Carlos I
de F{\'\i}sica Te{\'o}rica y Computacional. Universidad de Granada.E-18071, Granada, Spain}
\author{Adrián Roig}
\affiliation{Departamento de Electromagnetismo y F{\'\i}sica de la Materia and Instituto Carlos I
de F{\'\i}sica Te{\'o}rica y Computacional. Universidad de Granada.E-18071, Granada, Spain}
\author{Miguel A. Mu\~noz} 
\affiliation{Departamento de Electromagnetismo y F{\'\i}sica de la Materia and Instituto Carlos I
de F{\'\i}sica Te{\'o}rica y Computacional. Universidad de Granada.E-18071, Granada, Spain}

\date{\today}

\begin{abstract}
The idea that information-processing systems operate near criticality to enhance computational performance is supported by scaling signatures in brain activity. However, external signals raise the question of whether this behavior is intrinsic or input-driven.
We show that autocorrelated inputs and temporal resolution influence observed scaling exponents in simple neural models. We also demonstrate analytically that under subsampling, non-critical systems driven by independent autocorrelated signals can exhibit strong  signatures of apparent criticality.
To address these pitfalls, we develop a robust framework and apply it to pooled neural data, revealing resting-state brain activity at the population level is slightly sub-critical yet near-critical. Notably, the extracted critical exponents closely match predictions from a simple recurrent firing-rate model, supporting the emergence of near-critical dynamics from reverberant network activity, with potential implications for information processing and artificial intelligence.
\end{abstract}

\maketitle
Biological information-processing systems composed of many interacting units, such as neural networks, are believed to achieve significant computational advantages by operating close to a phase transition between different dynamical regimes \cite{ Plenz-review, munoz_colloquium_2018, chialvo_emergent_2010, Plenz-functional,   Breakspear-review, Kinouchi,  hesse_self-organized_2014, Beggs-book, Obyrne-review}. 
This idea has gained momentum in recent years, fueled both by increasingly precise high-throughput neural data revealing approximate scale invariance and criticality across spatiotemporal scales \cite{Meshulam-PRL, Morales-PNAS}, as well as by its active exploration in artificial neural networks and machine learning ---particularly within the framework of reservoir computing \cite{langton_computation_1990, BerNat, Luko, Boedecker}---  where multiple studies suggest that computational performance peaks when the network operates near criticality or ``the edge of instability" \cite{Morales-RC, Critical-RC-Wang}.

While the criticality framework is particularly appealing to theorists ---potentially serving as a unifying principle for biological information processing--- it has also sparked healthy criticism from various perspectives. A frequent concern is that these biological or artificial systems are not isolated but are constantly influenced by external signals. This raises the question of whether observed signatures of criticality reflect intrinsic critical-like reverberation within recurrent networks or are shaped by complex external inputs \cite{Latham}. For example, statistical signatures of criticality have been reported in neural population activity \cite{Mora-Bialek}, but similar patterns can also arise from models of independent neurons driven by structured latent inputs \cite{Nemenman1}. Likewise, phenomenological renormalization group (PRG) analyses have revealed scale invariance in large neuronal populations \cite{Kadanoff, Meshulam-PRL, Morales-PNAS}, though such behavior may also be derived from latent-variable models without genuine criticality \cite{Nemenman2}.
\begin{figure*}[tbh!]
    \centering
\includegraphics[width = 1.0\linewidth]{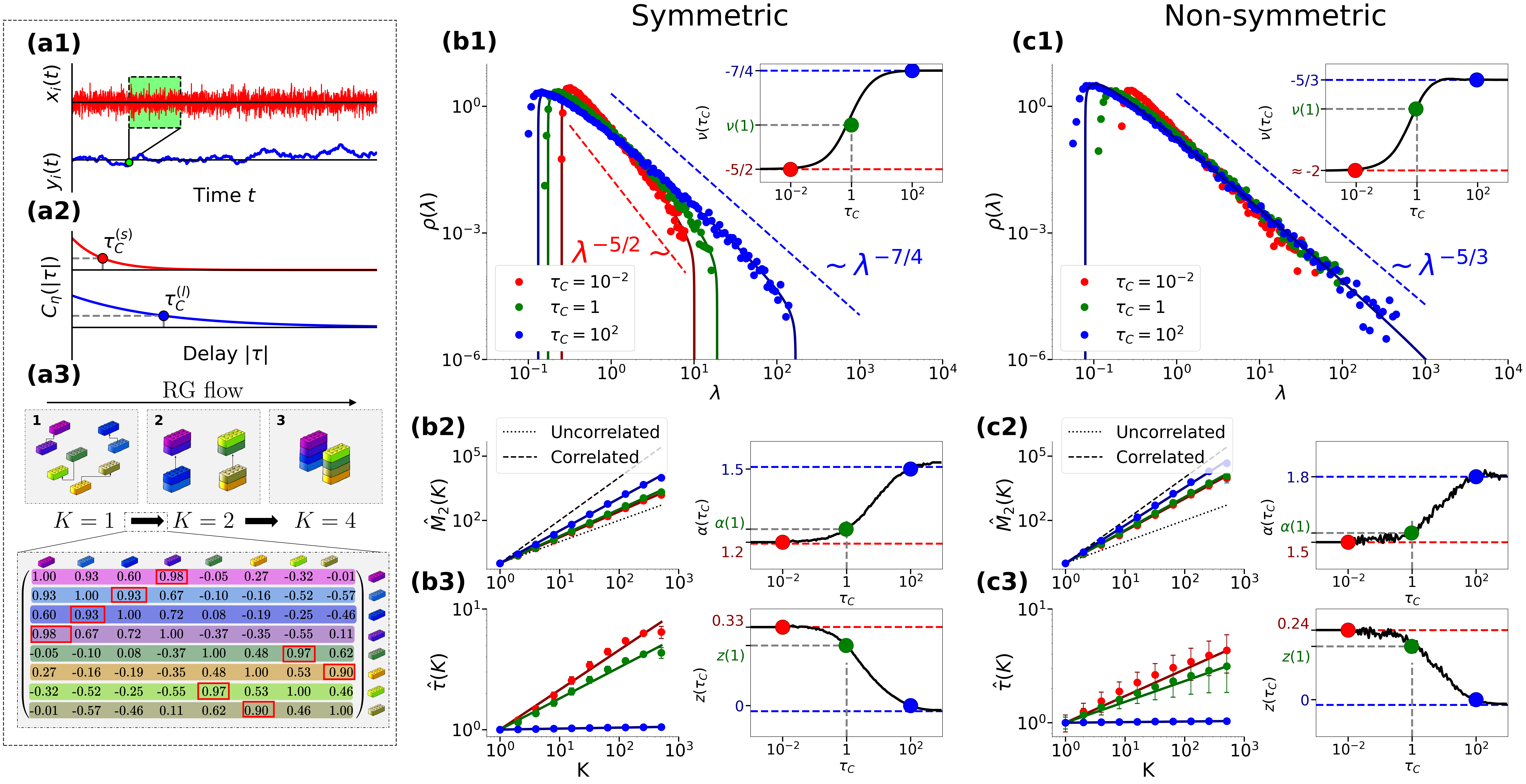}
\caption{\textbf{Time coarse-graining and time-correlated inputs both yield time-dependent critical exponents.} 
 \textbf{(a1)}. Illustration of a time series $ x_i(t) $ (red) and its coarse-grained counterpart $ y_i(t) $ (blue), which have a longer effective noise correlation time, as shown by the auto-correlation decay in  \textbf{(a2)}. 
 \textbf{(a3)}. Sketch of the PRG approach, iteratively merging maximally correlated units
 (e.g. marked in red boxes in the covariance matrix) into larger blocks.
 \textbf{(b1)/(c1)} Spectral density of the covariance matrix for symmetric (b1) and non-symmetric (c1) cases of the linear rate model for different correlation times $ \tau_c $. Dots: simulations; lines: theoretical predictions. Insets show the power-law scaling exponent $\nu$ vs. $\tau_c$.
\textbf{(b2)/(c2)} Scaling of the variance $\hat{M}_2(K)$ within clusters of size $K$ in between expected results for uncorrelated (dotted line) and fully correlated (dashed line) variables. Right plot: critical exponent $ \alpha $ vs. $ \tau_c $, for the symmetric (b2) and non-symmetric cases (c2).
\textbf{(b3)/(c3)} Scaling of the block intrinsic correlation time as a function of block size for the symmetric (b3) and non-symmetric cases (c3). Right plots: associated scaling exponent exponent $z$ interpolating between short- and long-time values}
    \label{fig. figure1}
\end{figure*}
From a quantitative viewpoint, Principal Component Analysis (PCA) \cite{jolliffe_principal_2002, Shlens, Shinn_phantom} offers a powerful approach to studying dynamical regimes in multivariate time series by reducing dimensionality via the covariance-matrix spectrum. Building on this, Hu and Sompolinsky (HS) \cite{HS} proposed estimating the distance to criticality by fitting the empirical spectrum to theoretical predictions from a linear firing-rate model with random connectivity driven by white noise. As the system approaches a phase transition to instability, this spectrum develops a broad power-law tail, reflecting a scale-invariant hierarchy of functional modes—an organization believed to optimize information representation \cite{Stringer-Nature}. This framework has revealed critical behavior in brain activity from microscopic recordings of individual neurons \cite{Stringer-Nature, Morales-PNAS}, further supported by alternative inference methods \cite{dahmen_second_2019, Morales-PNAS} and frequency-dependent covariance analyses of magnetoencephalography (MEG) data \cite{calvo_frequency-dependent_2024}. Together, these findings support the view that the healthy brain operates near the edge of instability. Yet, confirming this picture at the whole-brain level requires more evidence from functional magnetic resonance imaging (fMRI), a state-of-the-art and widely used technique, which faces challenges due to limited temporal resolution and strong temporal subsampling \cite{marcenko_distribution_1967}.
To advance in this direction, we report three key findings. First, we demonstrate that criticality signatures, such as scaling exponents, are non-universal because they are highly sensitive to temporal resolution; specifically, coarse-graining in time is effectively equivalent to introducing auto-correlations in external inputs, which in turn alters the scaling exponents. 
Second, we analytically show that under subsampling, the presence of strongly auto-correlated inputs can render strong apparent signatures of criticality, even in fully non-critical, uncoupled systems
Third, despite these challenges, the framework we propose enables the identification of genuine criticality in empirical whole-brain fMRI data at the population level, revealing robust and universal scaling exponents consistent with predictions from a simple theoretical model, thereby reinforcing the view that the brain operates near a critical regime.

\vspace{0.15cm}

Let us consider a set of time series $ x_i(t) $ from multiple measured units $ i = 1,2, \ldots,N$. 
PCA focuses on the structure of the ``short-time" covariance matrix: \footnote{For simplicity, we assume throughout this text that all empirical time series are standardized (zero mean and unit variance), allowing us to use the terms covariance and correlation interchangeably.}
\begin{equation} 
\label{eq. short-time}
C^0_{ij}(t) = \langle x_i(t) x_j(t) \rangle,
\end{equation}
where $i$ and $j$ index two units, and
$\langle \cdot \rangle$ denotes the average over measurements or realizations, typically estimated by time averaging assuming stationarity and ergodicity \cite{jolliffe_principal_2002, Shlens, Shinn_phantom}. 
However, it is often convenient to study a time-coarse-grained version of Eq. (\ref{eq. short-time}) using a time bin $\Delta T$:
\be
C_{ij}^{\Delta T} = \langle y_i(t) y_j(t) \rangle, \quad \text{with} \quad y_i(t) = \int_t^{t+\Delta T} \frac{x_i(s)}{\sqrt{\Delta T}} \, ds,
\label{binning}
\ee
i.e., the covariance matrix of the time-integrated variables $y_i(t)$, for $i = 1, 2, \ldots, N$ \cite{Pernice2011, Ocker2017}
(see Fig.\ref{fig. figure1}(a1)). Short-time windows capture fast dynamics but yield noisier estimates, while longer windows provide more stable results but may miss transient behaviors \cite{Averbeck,Meshulam-PRL,Morales-PNAS}. A common practice is to choose $\Delta T$ as large as possible, so that $C^{\infty}_{ij}$ effectively averages out temporal structure.

To gain analytical insight and following previous work \cite{sompolinsky_chaos_1988,HS}, we consider a firing-rate model on a random
recurrent neural network:
\begin{equation}
    \label{eq. firing rate model}
    \dot{x}_i(t) = -x_i(t) + g \sum_{j=1}^N W_{ij} S(x_j(t)) + \sigma \xi_i(t),
\end{equation}
where $x_i$ represents the firing rate of neuron $i$ ($i=1, \dots, N$), $g$ is the overall coupling strength, $W_{ij}$ is the synaptic weight from neuron $j$ to neuron $i$, $S(x)$ is a  response function (commonly $S(x) = \tanh(x)$), and $\sigma \xi_i(t)$ is the input signal, modeled as uncorrelated white noise with zero mean, unit variance, and amplitude $\sigma$. We analyze the case of random connectivity (RC) networks,  with independent Gaussian synaptic weights of zero mean, variance $\overline{W_{ij}^2} = 1/N$, and reciprocity $\overline{W_{ij} W_{ji}} = \gamma / N$ ($\gamma \in [-1,1]$, where $\gamma=1$ is symmetric and $\gamma=0$ non-symmetric). The overline denotes averaging over the synaptic-weight distribution. 
We focus on the linear case $S(x) = x$, with robustness against nonlinearities confirmed in the Supplemental Material \footnote{\setcounter{SIfootnote}{\value{footnote}}See Supplemental Material.}. The resulting set of linear equations describes a multivariate Ornstein-Uhlenbeck (OU) process \cite{uhlenbeck_theory_1930}, for which the spectrum of $C^\infty$ can be analytically determined for both general random networks ($\gamma \neq 1$) and networks with symmetric or reciprocal interactions ($\gamma = 1$) \cite{HS}. As shown by HS, when the overall coupling $g$ approaches the edge of instability, the spectrum broadens and develops a power-law tail, $\rho(\lambda) \sim \lambda^{\nu}$, with exponents $\nu = -5/3$ in the general case and $\nu = -7/4$ for reciprocal links (see Fig. \ref{fig. figure1},(b1,c1)). Can we generalize these results to generic coarse-grained time-series? 

By integrating the OU process over a time window of duration $\Delta T$ using Eq.(\ref{binning}), one finds that the effective equation for the coarse-grained variables (the $y$'s) is identical to that of the original variables (the $x$'s), with the crucial difference that the noise is no longer white but colored, that is, it exhibits temporal autocorrelations (see Fig.~\ref{fig. figure1}(a1,a2) and App.A). 
Building on this insight, we study the linear firing-rate model in which each neuron is now driven by an independent external input modeled as colored noise $\eta_i(t)$ with autocorrelation time $\tau_c$.
\begin{equation} \label{Eq. correlation OU}
    \langle \eta_i(t)\eta_j(s)\rangle = \delta_{ij}\frac{f(\tau_c)}{2} \exp\left(-\frac{|t-s|}{\tau_c}\right),
\end{equation}
where $f(\tau_c)$ is a normalization constant. As shown in App.A, we have derived an analytical expression (Eq.(\ref{eq. correlation short-to-long})) for the associated covariance matrix for any value of $\tau_c$ \cite{Liegeois}. 
Moreover, from this expression, we have been able to derive the spectral distribution for any coarse-graining level $\tau_c$ in the symmetric-coupling case ($\gamma = 1$) (Eq.\ref{eq. density}). As the coupling strength $g$ approaches the edge of instability, the spectral density develops a power-law tail with a $\tau_c$-dependent exponent $\nu(\tau_c)$, which in the symmetric case evolves smoothly from $\nu(\tau_c \to 0) = -5/2$ to $\nu(\tau_c \to \infty) = -7/4$, thus recovering the limiting results of HS \cite{HS} (Fig.\ref{fig. figure1}(b1,c1)). For the non-symmetric case ($\gamma = 0$), although an exact derivation is not available, numerical simulations show that $\nu(\tau_c)$ interpolates from approximately $-2$ to the known value $-5/3$ at large $\tau_c$ \cite{HS}. To our knowledge, the analytical dependence of scaling exponents on input autocorrelations $\tau_c$—equivalent to varying the coarse-graining $\Delta T$—has not been previously reported.

Similarly, we examined the effect of colored inputs on the scaling exponents of the phenomenological renormalization group (PRG) \cite{Meshulam-PRL}. Spatial coarse-graining involves merging units in pairs based on maximal covariance and iterating the process at various scales (see Fig.\ref{fig. figure1}(a3)). At each step, one measures the total variance inside each cluster of units of size $K$, $M_2(K)$, as well as the auto-correlation time $\tau(K)$. Both magnitudes are rescaled to $\hat{M_2}(K)=M_2(K)/M_2(0)$,  $\hat{\tau}(K)=\tau(K)/\tau(0)$ and represented against the cluster size $K$ in Fig.\ref{fig. figure1}(b2,b3,c2,c3) for both the symmetric and the non-symmetric networks. In all cases, these quantities exhibit non-trivial scaling $\hat{M_2}(K)\sim K^\alpha$ and $\hat{\tau}(K) \sim K^z$ when the system is set close to the edge of instability.
%($g=0.95/2$ and $g=0.95$, resp.).
The critical exponents $\alpha$ and $z$ vary continuously as functions of the input’s autocorrelation time $\tau_c$ \cite{Morales-PNAS} \footnote{Interestingly, the critical exponent $z$ converges to $0$ as $\tau_c\rightarrow\infty$ both in the symmetric and non-symmetric case, because the long-time window limit effectively averages away all the temporal structure.}.
In summary, these analyses reveal that the chosen time bin ---or equivalently, the temporal auto-correlation of noisy inputs--- can strongly influence the scaling behavior of recurrent networks near the edge of instability. 
This raises the question of whether auto-correlated inputs could induce even more drastic changes, potentially making a non-critical system appear critical especially in empirical data analyses where statistics are limited.
\begin{figure*}[tbh!]
    \centering
    \includegraphics[width = 1.0\linewidth]{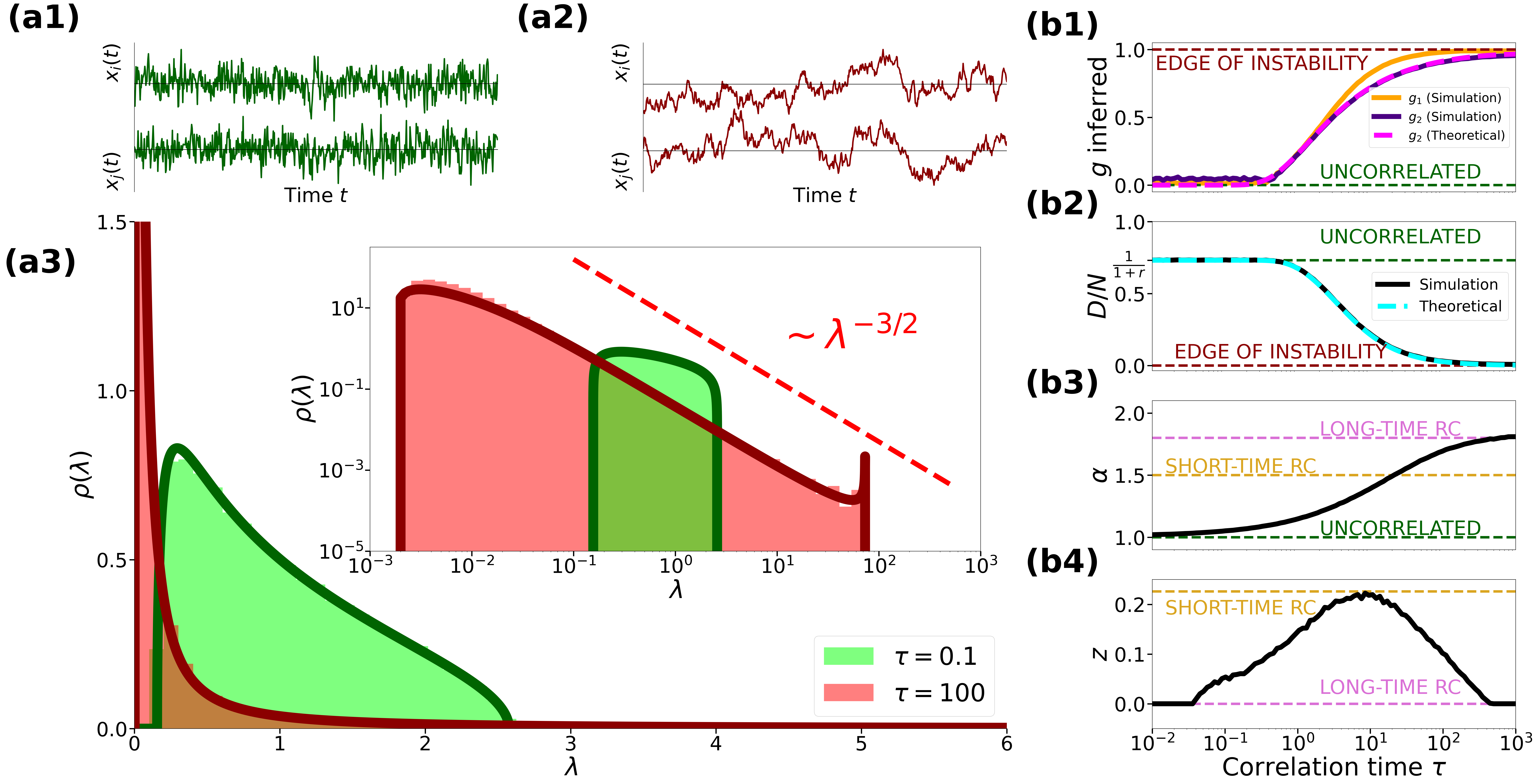}
    \caption{\textbf{Temporal correlations in independent units can induce critical-like behavior.} \textbf{(a1)} and \textbf{(a2)}. 
Time series of an Ornstein-Uhlenbeck process with short ($\tau = 0.1$) and long ($\tau = 100$) correlation times, respectively. 
\textbf{(a3)} Spectral density of the empirical covariance matrix ($N=183$ and $T=500$) for different $\tau$ values in natural (main) and logarithmic (inset) scales. Histograms: simulations; green line: Marchenko-Pastur distribution; red: our theoretical prediction for large $\tau$, showing a power-law tail ($\lambda^{-3/2}$) and a broad support.
\textbf{(b1)} Estimated distance to the edge of instability $ g $ vs. $\tau$, obtained via fitting to the Hu-Sompolinsky (HS) distribution, the second estimator $g_2$, together with our theoretical prediction for it.  
\textbf{(b2)} 
Dimensionality $D/N$ versus $\tau$ is shown with the dashed curve representing the theoretical prediction; in the uncorrelated limit, the Marchenko-Pastur dimension is $1/(1+r)$ where $r = N/T$.
\textbf{(b3)/(b4)} Scaling exponents $\alpha$ and $z$ vs. $\tau$, respectively, both revealing non-trivial behavior that can resemble recurrent network dynamics at criticality.}
    \label{fig. figure2}
\end{figure*}

\begin{figure*}[tbh!]
    \centering
    \includegraphics[width = 1.0\linewidth]{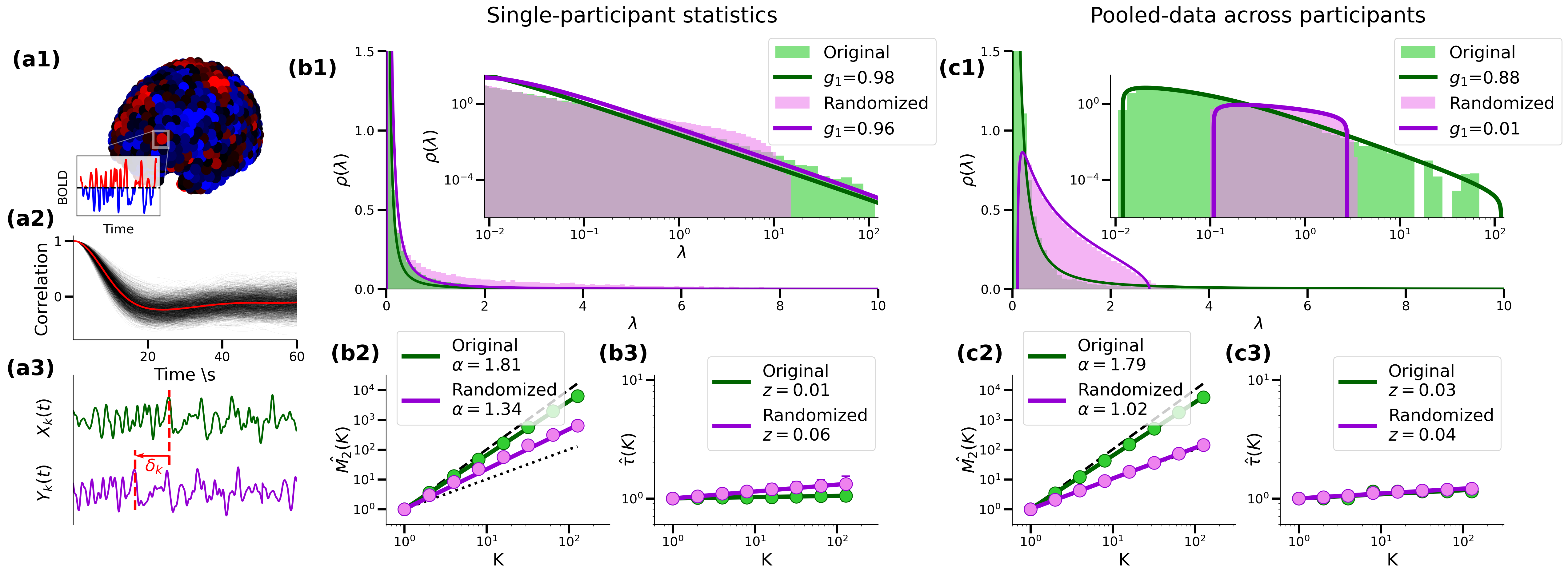}
    \caption{\textbf{Critical behavior in fMRI data is undetectable at the single-participant level but emerges at the collective (group) level.}\textbf{(a1)}. 3D brain image from fMRI, with BOLD signal above (red) and below (blue) average. 
\textbf{(a2)} Self-correlation across regions, showing consistently $15s$ correlation time. 
\textbf{(a3)} Time-series before (green) and after (purple) randomization for a given subject, which preserves correlation times but removes spatial correlations. 
\textbf{(b1)/(c1)} Spectral density of the empirical covariance matrix for single participants (b1) and pooled data (c1), before (green) and after (purple) randomization (insets: logarithmic scale). Fits to the Hu-Sompolinsky distribution give near-critical values $ g_1 \approx 0.98 $ (original) and $ g_1 \approx 0.96 $ (randomized) for (b1), but a significant drop from $ g_1 \approx 0.88 $ to $ g_1 \approx 0.01 $ in (c1). 
\textbf{(b2)/(c2),(b3)/(c3)} Renormalization group analyses of the activity variance $\hat{M}_2(K)$ and correlation time $\hat{\tau}(K)$, showing non-trivial scaling in (b2), for individual subjects, even after randomization, but a strong reduction in (c2), for pooled data, approaching the uncorrelated case. The critical exponent $ \nu $ also decreases significantly in (c3).}
    \label{fig. figure3}
\end{figure*}
To explore this, we analyze a set of $N$ statistically independent neurons modeled by the uncoupled linear firing-rate system ---i.e. Eq. (\ref{eq. firing rate model}) with $g = 0$--- driven by noise with a tunable autocorrelation time $\tau$ (Fig. \ref{fig. figure2}a1/a2). We focus on a regime of limited sampling due to a finite simulation time $T$, so that the units-to-sample ratio $r = N/T$ remains far from the asymptotic limit $r \to 0$.
As detailed in App.B, the spectral density of the empirical covariance matrix can be computed by solving a quartic equation derived from random matrix theory \cite{burda_spectral_2005}. In the limit of short input autocorrelation times ($\tau \to 0$), the solution recovers the well-known Marchenko–Pastur law \cite{marcenko_distribution_1967, Bouchaud}, while for large $\tau$, we approximate the solution by neglecting higher-order terms, enabling an analytic
expression for the spectral density
 \footnotemark[2]:
\begin{equation}
\label{eq.distribution+correlations}
    \rho(\lambda)=\frac{1}{\pi\lambda}\frac{\sqrt{(\lambda_--\lambda)(\lambda-\lambda_+)}}{(\lambda_--\lambda)(\lambda-\lambda_+)+1},
\end{equation}
where $\lambda_\pm=r(c\pm\sqrt{c^2-1})$ with $c=\coth(1/\tau)$, and support $\lambda\in[\lambda_-, \lambda_+]$.  
Remarkably, as $\tau \to \infty$, the spectral support broadens ($\lambda_- \to 0$, $\lambda_+ \to \infty$), and the distribution develops a  power-law tail with exponent $\nu = -3/2$ (Fig.\ref{fig. figure2}(a3)), mimicking the hallmark signature of a genuinely critical system, despite the complete absence of interactions.
\footnote{Observe that, when $c$ is large enough, $\lambda_+\sim rc$; given that $c\sim \tau$ for large values of $\tau$, one can conclude that $\lambda_+\sim N/(T/\tau)$, meaning that the upper edge of the support of this distribution is controlled by the ratio of the correlation time to the duration of the time-series.}
Actually, one can estimate the apparent distance to criticality, $ g_c = 1 $, as a function of $ \tau $ using two established methods \cite{Morales-PNAS}: (i) $ g_1 $, obtained by fitting the empirical covariance spectrum to the HS distribution \cite{HS}, and (ii) $ g_2 $, derived from the relative dispersion of covariance matrix entries \cite{dahmen_second_2019} (following \cite{dahmen_second_2019}, we were also able to derive an analytical expression for $g_2$ in the case of independent units exposed to auto-correlated signals in terms of $\tau$, $N$, and $T$; see App.C). Results shown in Fig.\ref{fig. figure2}(b1) reveal strong consistency between both estimators and with our theoretical predictions as a function of $\tau$. Moreover, Fig.\ref{fig. figure2}(b2) shows the normalized participation dimension $D/N$, which quantifies the effective number of independent modes contributing to the system's dynamics, versus $\tau$. Simulation results (black solid line) and theoretical predictions (blue dashed line) perfectly match, revealing a dimensionality collapse ---as typically observed at criticality--- as the correlation time increases. Similarly, the PRG also reveals non-trivial scaling exponents when $\tau\rightarrow \infty$, as shown in Fig.\ref{fig. figure2}(b3,b4). This underscores that blind estimations of the dynamical regime with limited sampling can mistakenly suggest near-critical behavior, as unstructured but auto-correlated latent variables may create the illusion of criticality even in uncoupled neurons. 

To address this, a principled test involves randomly time-shifting each neuron's activity trace: this disrupts genuine inter-neuronal correlations—eliminating true criticality—while preserving individual autocorrelations that drive spurious effects. This makes it especially valuable for empirical analyses where such distinctions are essential. While time-shift controls have been effectively applied to individual spiking activity \cite{Morales-PNAS} and MEG data \cite{calvo_frequency-dependent_2024}, greater caution is needed in whole-brain studies using functional magnetic resonance imaging (fMRI), where temporal resolution is low and latent autocorrelations can be significantly large \footnote{fMRI relies on the blood-oxygen-level-dependent (BOLD) signal, reflecting fluctuations in neural activity across cortical regions (see Fig. \ref{fig. figure3}a1)).}. Although fMRI provides high spatial resolution, its temporal resolution is limited to approximately 1 Hz over 10–15 minutes, resulting in $T \sim 10^3$ time points. This limits meaningful analyses to fewer than $10^3$ regions of interest (ROIs), ensuring $T > N$, particularly when auto-correlations are strong. To illustrate potential pitfalls, we analyzed resting-state fMRI data from the LEMON dataset \cite{LEMON}, which includes $136$ participants, each with $T = 652$ time points sampled every 1.4 seconds. To maximize the units-to-samples ratio $r$, we used a reduced parcellation with $N = 183$ ROIs.
For each participant, we computed the covariance-matrix spectrum, which, as shown in Fig.\ref{fig. figure3}(b1), exhibits a clear power-law tail. Fitting this spectrum with the HS model yielded an effective coupling strength of $g_1 \approx 0.98 \pm 0.01$, suggestive of near-critical dynamics. PRG analyses further revealed non-trivial scaling exponents (Fig.\ref{fig. figure3}(b2,b3)). However, individual time series exhibited long auto-correlation times ($\approx$15s, Fig.\ref{fig. figure3}(a2)), raising concerns that the apparent criticality may be spurious.

To check this,
we implemented the time-shift randomization test mentioned before, which preserves the temporal structure but disrupts correlations between ROIs (see Fig.\ref{fig. figure3}(a3)). After applying this transformation, the spectral density still exhibited near-critical behavior, with $g_1 \approx 0.96 \pm 0.10$ (Fig.\ref{fig. figure3}(b1)), and non-trivial scaling exponents remained in the PRG analyses (Fig.\ref{fig. figure3}(b2/b3)), suggesting that the observed signatures of criticality likely result from strong under-sampling in single-participant data.

To mitigate this issue, we performed a pooled analysis by concatenating data from all participants, treating them as separate sessions of a single subject. This collective approach yields a spectral density exhibiting slightly sub-critical behavior, yet remaining close to criticality ($g_1 \approx 0.88$). In contrast, when each participant’s time series are randomized (Fig.~\ref{fig. figure3}(c1)), the coupling drops to $g_1 = 0.01$, indicating uncorrelated dynamics—consistent with expected loss of meaningful structure. The scaling exponent $\alpha$ shifted from $\approx 1.79$ (compatible with the firing-rate model’s long-time results) to $\sim 1.02$, consistent with uncorrelated variables.

These findings show that, after addressing spurious correlations due to limited sub-sampling and pooling across participants, the near-critical behavior observed in the spectral density reflects robust, slightly sub-critical dynamics at a collective level, rather than artifacts arising from data limitations or long correlation times. Remarkably, the critical exponents observed in the empirical data closely match those predicted by the linear firing-rate model, suggesting universal scaling behavior.
%($\nu = -5/3$, $\alpha = 7/4$, and $z \approx 0$).

We hope the insights and methodologies presented here will inspire advances across fields—from machine learning to microbial ecology—where claims of near-critical dynamics continue to emerge, and guide the development of critical reservoir computing strategies.

\vspace{0.15cm}
{\bf{Acknowledgments:}} 
We acknowledge the Spanish Ministry %of Research and Innovation 
and Agencia Estatal de Investigación (AEI), MICIN/AEI/10.13039/501100011033, for financial support, Project PID2023-149174NB-I00 funded also by ERDF/EU. We thank G.B. Morales and S. di Santo for insightful discussions, feedback, and collaboration.

\newpage

\section*{End Matter}

\subsection{Appendix A: From short to long-time window covariances}
Consider the stochastic differential equation given by Eq.(\ref{eq. firing rate model}) in the case $S(x)=x$, i.e. the linear firing-rate model. Its associated short-time window covariance matrix does not admit a closed solution in terms of the coupling matrix $W$, but is rather  given as the solution of the Lyapunov equation \cite{barnett_ch_1971}:
\begin{equation}  \label{Eq: Lyapunov}
    (I - gW) C^0 + C^0 (I - gW)^T = \sigma^2 I,
\end{equation} 
where $I$ is the identity matrix. In contrast, the long-time window covariance matrix can be explicitly expressed in terms of $W$ as \cite{HS, calvo_frequency-dependent_2024}:  
\begin{equation}  \label{Eq. long-time window covariance}
    C^{\infty} = \sigma^2 (I - gW)^{-1} (I - gW)^{-T}. 
\end{equation}  
Integrating both sides of Eq.(\ref{eq. firing rate model}) over a time window of size $\Delta T$ yields a new process for the coarse grained variables, $y_i(t)$,  Eq.(2), which is identical to Eq.(\ref{eq. firing rate model}) but  driven by a time-correlated Gaussian noises $\eta_i(t)$, with auto-correlations obeying
\begin{equation}
    \langle \eta_i(t)\eta_j(s)\rangle = \sigma^2\frac{\delta_{ij}}{\Delta T}\int_t^{t+\Delta T}\int_s^{s+\Delta T} \delta(u -v)\, dudv,
\end{equation}
or equivalently:
\begin{equation}
\label{eq: tent correlation}
  \langle \eta_i(t)\eta_j(s)\rangle =
\begin{cases}
\sigma^2\left(1-\dfrac{|t-s|}{\Delta T}\right) \,\delta_{ij}, & \text{if } |t-s| \le \Delta T,\\[1mm]
0, & \text{if } |t-s| > \Delta T.
\end{cases}  
\end{equation}
which is the tent function with height $\sigma^2$ and width $\Delta T$. It is easy to prove that the covariance matrix $C^{\Delta T}$ for 
time windows $\Delta T$ is:
\begin{equation}
\label{eq. C delta T}
    C^{\Delta T} = C^{\infty}-\left( B(\Delta T)C^0+C^0 B(\Delta T)^T \right), 
\end{equation}  
where 
\begin{equation} 
\label{eq. B Delta T}
    B(\Delta T) = \frac{\left(Id-e^{-A\Delta T}\right)}{\Delta T}\frac{1}{A^2},
\end{equation}  
and $A=(I-gW)$. Observe that  Eq.(\ref{eq. C delta T}) is an interpolation between the long and the short-time window limits for $\Delta T\to \infty$
 and $\Delta T\to 0$, respectively.

To make further analytical progress
and be able to explicitly derive the shape of the spectrum as a function of input auto-correlations, we considered a simpler form of noise correlations; i.e. that of Eq.(\ref{Eq. correlation OU}).
 After some algebra, it is easy to derive the following expression for the covariance matrix of the process $y_i(t)$, driven by $\eta$, which we will denote as $C_Y$:
\begin{eqnarray} \label{eq. correlation short-to-long}
C_Y(\tau_c)&=& \frac{\sigma^2 f(\tau_c)}{2}\left(A+1/\tau_c\right)^{-1}\left( A-1/\tau_c \right)^{-T}  \nonumber\\
&-&\frac{f(\tau_c)}{\tau_c}C^0\left(A+1/\tau_c\right)^{-T}\left( A-1/\tau_c \right)^{-T}.
\end{eqnarray} 
Choosing $f(\tau) = 1 + 1/\tau$ guarantees the limits $C_Y(\infty) = \sigma^2 A^{-1}A^{-T}$ and $C_Y(0) = C^0$, thus recovering the long and short-time window covariances, Eq.(\ref{Eq. long-time window covariance}) and Eq.(\ref{eq. short-time}), of the original process when $\Delta T = \tau$.

Observe that Eq.(\ref{eq. correlation short-to-long}) is simpler in that it does not involve any matrix exponentials. In fact, we can derive the spectral distribution of $C_Y$ for symmetric coupling matrices ($\gamma = 1$) for any value of $\tau_c$ using the theorem of change of variables, which yields: 
\begin{widetext} 
\begin{equation} \label{eq. density} 
\rho(\lambda)=\frac{1}{2\pi \sigma^2 g^2} \frac{1+\tau_c}{2\lambda^2\sqrt{1+\frac{2\tau_c(\tau_c+1)}{\lambda}}} \sqrt{4g^2 -\left( \frac{1}{2\tau_c}\left( 1-\sqrt{1+\frac{2\tau_c(\tau_c+1)}{\lambda}} \right) +1\right)^2},
\end{equation} 
\end{widetext} 
with support $\lambda \in (\lambda_-, \lambda_+)$, where 
\begin{equation} 
\lambda_{\pm}=\frac{\sigma^2}{2(1\pm 2g)(\tau_c+1)}\left( \frac{\tau_c}{2(1\pm 2g)}+1 \right),
\end{equation} and $g \in [0, 1/2]$.

\subsection{Appendix B: Spectral density for auto-correlated time series}
Let $X$ be an empirical time-series of an $N$-dimensional stochastic process sampled at $T>N$ different time-points.
We denote the empirical covariance matrix as:
\begin{equation}
    \tilde{C}=\frac{1}{T}XX^T
\end{equation}
Since $X$ is a sample of a stochastic process, its entries are random variables with correlations given by:
\begin{equation}
    \langle X_{i\alpha}X_{j\beta}\rangle = C_{ij}A_{\alpha\beta}, ~i,j\in\{1,...N\},\;\alpha,\beta\in\{1,...T\},
\end{equation}
where $C$ is the true correlation matrix across units, and $A$ captures the temporal structure.
 For example, if $C_{ij}=\delta_{ij}$, $A_{\alpha \beta}=\delta_{\alpha\beta}$, the ensemble of all possible empirical matrices $\tilde{C}$ is known as the Wishart ensemble \cite{burda_spectral_2005}. Even when the stochastic process $X$ is uncorrelated, the spectral properties of the Wishart ensemble may deviate significantly from the expected result (which should be a Dirac's delta placed at $1$) if T is not sufficiently large, as proved by  Marchenko and Pastur  \cite{marcenko_distribution_1967}. Here, following \cite{burda_spectral_2005} we look at the spectral properties of this ensemble in the case in which:
\begin{equation}
    C_{ij}=\delta_{ij},\;\; A_{\alpha\beta}=\exp\left( -|\alpha-\beta|/\tau\right),
    \label{A}
\end{equation}
where $\tau$ stands for the correlation time of the series. 
As shown in \cite{burda_spectral_2005}, the spectral density of  $\tilde{C}$ is a solution of the quartic equation:
\begin{equation}
\label{eq: quartic}
    x^4+2x^3(r-cz)+x^2(r^2-2crz+z^2-1)-2rz-r^2=0,
\end{equation}
where $r=N/T$, $x(z)=rm(z)$, $m(z)$ is the moment generating function of $\rho(\lambda)$, and $c=\coth(1/\tau)$. 
The spectral density $\rho(\lambda)$ is encoded in the function $m(z)$ and can be retrieved via the transformation \cite{burda_spectral_2005}:
\begin{equation}
\label{eq: Spectral transform}
    \rho(\lambda)=-\lim_{b\to 0^+} \left( \frac{1+m(\lambda+ib)}{\pi (\lambda+ib)} \right).
\end{equation}
Eq.(\ref{eq: quartic}) can, in principle, be solved via the Ferrari method, but the resulting expressions are analytically intractable. 
Instead, we use a perturbative approximation that neglects cubic and quartic terms, justified in the $\tau \to \infty$ limit, where the spectral density vanishes in the tails. Solving the resulting quadratic equation gives:
\begin{equation}
    m(z)=\frac{1\pm\sqrt{r^2-2crz+z^2}}{-1+r^2-2crz+z^2},
\end{equation}
and using Eq.(\ref{eq: Spectral transform}) we obtain:
\begin{equation}
    \rho(\lambda)=-\frac{1}{\pi\lambda}\frac{\sqrt{|\lambda^2 -2cr\lambda+r^2|}}{\lambda^2 -2cr\lambda+r^2-1}
\end{equation}
and the support is the region where $\Delta=\lambda^2 -2cr\lambda+r^2<0$, for which there is a an imaginary solution for $m(z)$. Solving for $\Delta=0$ gives the limits of the support $\lambda\in[\lambda_-,\lambda_+]$, where $\lambda_\pm = r(c\pm\sqrt{c^2-1})$,
in agreement with Eq.(\ref{eq.distribution+correlations}) in the main text.
As $\tau \to \infty$, $\rho(\lambda) \sim \lambda^{-3/2}$, revealing a power-law decay, even if the system is uncoupled.

\subsection{Appendix C: Alternative estimators of criticality}

An alternative to the HS method to infer the  distance to instability was introduced by Dahmen et al. in \cite{dahmen_second_2019}. 
This estimator quantifies the dispersion of pairwise covariances normalized by the squared mean of the diagonal (auto-correlation) 
$ \Delta^2 = \langle (C^\infty_{i,j \neq i})^2 \rangle / \langle C^\infty_{i,i} \rangle^2$
where the averages are taken over all pairs $(i, j)$,
and exploits the fact that the histogram of cross-correlations broadens as the system approaches criticality \cite{dahmen_second_2019}. 
\begin{comment}
The estimator quantifies the dispersion of pairwise covariances, normalized by the squared mean of the diagonal (auto-correlation) terms: 
\begin{equation} \label{eq. Delta} 
\Delta^2 = \frac{\langle (C^\infty_{i,j \neq i})^2 \rangle}{\langle C^\infty_{i,i} \rangle^2}, 
\end{equation} 
where the averages are taken over all node pairs $(i, j)$. 
\end{comment}
For the firing-rate model (Eq.~\ref{eq. firing rate model}) with a fully non-symmetric Gaussian matrix of zero mean and unit variance, the coupling strength $g$ can be expressed in terms of $\Delta^2$ as: \begin{equation} \label{eq. Dahmen estimator}
g = \sqrt{1 - \sqrt{\frac{1}{1 + N\Delta^2}}}. \end{equation} 
Under the assumption of linear firing-rate dynamics, an effective $g$ can thus be inferred from measurements of $\Delta$.
%where $N$ is the network size. 
\begin{comment}
Since correlations diverge at criticality, $\Delta$ also diverges, making $g$ an effective estimator of the distance to criticality. Under the assumption of linear firing-rate dynamics, an effective $g$ can thus be inferred from measurements of $\Delta$.
\end{comment}
We have extended this result to
deal with the correlated Wishart ensemble
as defined above.
%we derived an expression for this estimator in terms of the 
%correlation time $\tau$ and the units-to-samples ratio $r$. Indeed, let us start with the calculation of:
The average of the square of the pairwise covariances can be written as
\begin{equation}
    \langle C_{i,j}^2\rangle = \frac{1}{N^2}\sum_{i\neq j}\left( \frac{1}{T}\sum_{\alpha} X_i(t_\alpha)X_j(t_\alpha) \right)^2,
\end{equation}
and exchanging the order of the sums:
\begin{equation}
    \frac{1}{T^2}\sum_{\alpha\beta}\left(\langle X_i(t_\alpha)X_i(t_\beta)\rangle\right)^2= \frac{1}{T^2}\sum_{\alpha\beta} A_{\alpha\beta}^2.
\end{equation}
Using Eq.(\ref{A}), the expression for the sum of a geometric series, and assuming that $T,\:N>>1$, we obtain:
\begin{equation}
    N\Delta^2\approx r\left( {2}\left(\frac{1-q^T}{1-q}\right)-1 \right).
\end{equation}
Plugging this into Eq.(\ref{eq. Dahmen estimator}), finally, defining $q = \exp(-2/\tau)$ yields
an analytical prediction:
\begin{equation} \label{eq. Dahmen estimator MP} 
g = \sqrt{1 - \sqrt{\frac{1}{1 + {2}r\left(\frac{1 - q^T}{1 - q} - 1\right)}}}.
\end{equation} 

Finally, another estimator of criticality is the inverse participation ratio $D$, 
\begin{equation}\label{DimensionMOMENTS} 
D/N = \langle \lambda \rangle^2 /{\langle\lambda^2\rangle} \ , \end{equation} 
%where $E[x]$ indicates the expectation of the random variable $x$.
that measures effective dimensionality and vanishes near the edge of instability in firing-rate models
\cite{dahmen_second_2019,clark_dimension_2023}.
For the correlated Wishart ensemble, $D$ can be computed analytically from the known first and second moments of the eigenvalue distribution (see above and \cite{burda_spectral_2005}):
\begin{equation} \label{eq. Dimension MP}
%\frac{D}{N} = \frac{1}{1 + r + 2r\frac{c^2}{1 - c^2}}.
D/N= [1 + r + 2r c^2 /(1 - c^2]^{-1}.
\end{equation}
In particular, when $\tau\to 0$, one recovers the dimension for the Marchenko-Pastur distribution, which is $1/(1+r)$ and as $\tau \rightarrow \infty$ 
the dimension collapses as in a critical process.

\end{document}